\documentclass[12pt]{iopart} 

\usepackage{graphicx}
\usepackage{dcolumn} 
\usepackage{bm} 
\usepackage{epsfig}
\usepackage{subfigure} 
\usepackage{color}

\begin{document}
\title{Ultrarelativistic Transport Coefficients in Two Dimensions}

\author{M. Mendoza} \ead{mmendoza@ethz.ch} \address{ ETH Z\"urich,
  Computational Physics for Engineering Materials, Institute for
  Building Materials, Schafmattstrasse 6, HIF, CH-8093 Z\"urich
  (Switzerland)}

\author{I. Karlin} \ead{karlin@lav.mavt.ethz.ch} \address{ETH
  Z\"urich, Department of Mechanical and Process Engineering,
  Sonneggstrasse 3, ML K 20, CH-8092 Z\"urich (Switzerland)}

\author{S. Succi} \ead{succi@iac.cnr.it} \address{Istituto per le
  Applicazioni del Calcolo C.N.R., Via dei Taurini, 19 00185, Rome
  (Italy), and Freiburg Institute for Advanced Studies, Albertstrasse,
  19, D-79104, Freiburg, (Germany)}

\author{H. J. Herrmann}\ead{hjherrmann@ethz.ch} \address{ ETH
  Z\"urich, Computational Physics for Engineering Materials, Institute
  for Building Materials, Schafmattstrasse 6, HIF, CH-8093 Z\"urich
  (Switzerland), Departamento de F\'isica, Universidade Federal do
  Cear\'a, Campus do Pici, 60455-760 Fortaleza, Cear\'a, (Brazil)}

\date{\today}
\begin{abstract}
  We compute the shear and bulk viscosities, as well as the thermal
  conductivity of an ultrarelativistic fluid obeying the relativistic
  Boltzmann equation in $2+1$ space-time dimensions.  The relativistic
  Boltzmann equation is taken in the single relaxation time
  approximation, based on two approaches, the first, due to Marle and
  using the Eckart decomposition, and the second, proposed by Anderson
  and Witting and using the Landau-Lifshitz decomposition. In both
  cases, the local equilibrium is given by a Maxwell-J\"uttner
  distribution.  It is shown that, apart from slightly different
  numerical prefactors, the two models lead to a different dependence
  of the transport coefficients on the fluid temperature, quadratic
  and linear, for the case of Marle and Anderson-Witting,
  respectively. However, by modifying the Marle model according to the
  prescriptions given in Ref.~\cite{Takamoto}, it is found that the
  temperature dependence becomes the same as for the Anderson-Witting
  model.
\end{abstract}


\maketitle

\section{Introduction}

The study of the transport properties of 2D relativistic fluids from
the standpoint of kinetic theory is an important topic, still awaiting
a complete systematization.  Kremer and Devecchi \cite{kremer2D}
calculated the bulk viscosity of a two dimensional relativistic gas
using the Anderson-Witting collision operator and the Chapman-Enskog
expansion \cite{Anderson, RelaBoltEqua, momentclosure}, but did not
investigate the shear viscosity and thermal conductivity.

In this paper, we compute the transport coefficients, namely the bulk
and shear viscosities and the thermal conductivity, by using two
single relaxation time models (also called model equations). The first
one, proposed by Marle \cite{MarleModel}, is appropriate for mildly
relativistic fluids with moderate values of the Lorentz factor,
$\gamma < 2$.  The second one, by Anderson and Witting
\cite{Anderson}, on the other hand, can deal with significantly larger
Lorentz factors. The Marle model, as it was initially proposed in
Ref.~\cite{MarleModel}, is not appropriate to describe a gas of
ultrarelativistic particles, due to the fact that it implies an
infinite relaxation time in the limit where the mass of the particles
becomes zero, thus leading to divergent transport coefficients.
However, Takamoto et al. \cite{Takamoto} proposed a modified Marle
model, whereby the relaxation time of the Boltzmann equation is
redefined in such a way as to regulate the aforementioned infinities.
In addition, it is known that by promoting the relaxation time to the
status of a dynamic field, it is possible to describe complex flows
far from equilibrium, such as they occur in turbulence \cite{science}.
Therefore, this single relaxation time model will also be included in
the present study of the transport coefficients. In general, model
equations do not give the same transport coefficients as the ones
obtained from the full Boltzmann equation. However, it was proven in
the Ref.~\cite{chapmanmoments} that the methods of Chapman-Enskog and
Grad lead to the same approximations to transport coefficients, when
polynomial expansions of the distribution function in the peculiar
velocity are performed.

 In both cases, we use the moment expansion of the non-equilibrium
 distribution, similar to the fourteen fields\cite{RelaBoltEqua,
   otherkinetic3D, momentclosure} in the three-dimensional case.  To
 the best of our knowledge, this task has never been undertaken before
 for the case of two spatial dimensions.  This is all but an academic
 exercise, since two-dimensional relativistic flows arise in many
 areas of modern physics, say, cosmology, e.g. in galaxy formation
 from fluctuations in the early universe \cite{galaxyformation}, as
 well as in high-energy nuclear physics, e.g energetic heavy ions
 collisions \cite{heavyheavy}.  Two-dimensional ultrarelativistic
 fluids received a further boost of popularity in 2004, with the
 discovery of the gapless semiconductor graphene
 \cite{natletter,Geim1}.  This consists of literally a single carbon
 monolayer and represents the first instance of a truly
 two-dimensional material (the ``ultimate flatland''\cite{PhysToday}),
 where electrons move like massless chiral particles, whose dynamics
 is governed by the Dirac equation, with the Fermi velocity playing
 the role of the speed of light in relativity
 \cite{CastroNeto09,Peres10}. However, the calculation of the
 transport coefficients is more general and can be extended to any
 statistical system of quasi-particles governed by relativistic
 Boltzmann-like equations, i.e. it might apply to a whole class of
 systems where physical signals are forced to move close to the their
 ultimate limiting speed \cite{SR2012}.

 The results of this paper are restricted by the range of
 applicability of the Boltzmann equation to (quasi) two-dimensional
 systems. It is well known that linearizing hydrodynamics in two
 dimensions leads to divergent transport coefficients, both in
 classical and relativistic systems \cite{transanomalous,
   transanomalous2}. However, as long as the Boltzmann equation
 provides a useful semi-phenomenological approximation to transport
 phenomena, as for example evidenced by the use of the Boltzmann
 equation in quantum transport \cite{reviewgrap}, results of our
 computations remain valid.

 We wish to emphasize that the main goal of the present paper is to
 derive the transport coefficients for $2+1$ dimensional relativistic
 fluids, {\it out of prescribed relaxation times}.  In the
 non-relativistic case, this task is pretty straightforward, since in
 an absolute reference frame, there is no ambiguity as to the
 definition of the macroscopic observables (kinetic moments) in terms
 of the Boltzmann distribution.  In the relativistic case, on the
 other hand, this correspondence, i.e.  the projection from the
 kinetic to the hydrodynamic space, is much less direct and requires
 careful consideration.  Besides its theoretical interest, the
 practical target of this work is to provide operational input for
 lattice formulations of the Boltzmann equation, which have recently
 shown major potential for the numerical simulation of a broad class
 of relativistic flows across scales, from astrophysical flows, all
 the way down to quark-gluon plasmas \cite{LBE1,LBE2,rlbPRL,rlbhupp},
 including turbulent phenomena in the two-dimensional electronic gas
 in graphene \cite{turbPRL}.

 % As an additional advantage of having the transport coefficients is
 % that their implementation in numerical methods based on lattice
 % Boltzmann models \cite{LBE1, LBE2} becomes easier and therefore,
 % the study of complex systems where analytical solutions does not
 % exist, can be easily and efficiently studied numerically.

\section{Non-Equilibrium Distribution}

The single relaxation time Boltzmann equation for the Minkowski
metric, $\eta^{\alpha \beta}$, can be written as \cite{RelaBoltEqua}
\begin{equation}\label{Boltzmann:eqM}
  p^\mu \partial_\mu f = -\frac{mc}{\tau_M}( f - f^{\rm eq} ) \quad ,
\end{equation}
for the case of the Marle model \cite{MarleModel}, and as
\begin{equation}\label{Boltzmann:eqAW}
  p^\mu \partial_\mu f = -\frac{p^\mu U_\mu}{c^2 \tau}( f - f^{\rm eq} ) \quad ,
\end{equation}
for the case of Anderson and Witting model \cite{Anderson}. Here, $m$
is the mass of the particles, $c$ the speed of light, $k_B$ the
Boltzmann constant, $f$ the probability distribution function (which
can denote any scalar field in phase space), $\tau$ the single
relaxation time for the Anderson-Witting model, and $\tau_M$ the
respective one for the case of the Marle model. The 3-momentum is
denoted by $p^\mu = (p^0, \vec{p})$, and the macroscopic 3-velocity by
$U^\mu$.  Greek indices run from $0$ to $2$, being $0$ the temporal
component, and we have adopted the Einstein notation (repeated indexes
are summed). For the purpose of this work, we are using the signature
$(+,-,-)$. The equilibrium distribution $f^{\rm eq}$ is given by
\cite{RelaBoltEqua}
\begin{equation}\label{MJ:eq}
  f^{\rm eq} = A(n,T) \exp(-p_\mu U^\mu/k_B T) \quad , 
\end{equation}
where $A(n, T)$ is a normalization constant that depends on the
temperature and the number of particles density $n$. In this work, we
will study the ultrarelativistic regime, which is characterized by
$\xi \equiv mc^2/k_B T \ll 1$. From now on, we will use natural units,
$m = c = k_B = 1$, and the following notation:
\begin{eqnarray}
  \Delta^{\alpha \beta} &= \eta^{\alpha \beta} - U^\alpha U ^\beta
  \quad , \nonumber \\
  T^{(\alpha \beta)} &= \frac{1}{2} ( \Delta^\alpha_\gamma
  \Delta^\beta_\delta + \Delta^\beta_\gamma \Delta^\alpha_\delta  )
  T^{\gamma \delta}
  \quad , \nonumber \\
  T^{<\alpha \beta>} &= T^{(\alpha \beta)} - \frac{1}{2}
  \Delta^{\alpha \beta} \Delta_{\gamma \delta}
  T^{(\gamma \delta)}
  \quad , \nonumber \\
  \nabla^\alpha &= \Delta^{\alpha \beta} \partial_\beta \quad .
\end{eqnarray}

In order to identify the physical meaning of the different terms in
the balance and transport equations, it is useful to introduce
decompositions of these terms with respect to orthogonal
quantities. Note that $\Delta^{\alpha \beta}$ and $U^\alpha$ are
orthogonal quantities, $\Delta^{\alpha \beta} U_\beta = 0$, so that
any 3-vector can be decomposed into this orthogonal basis.  We begin
with the Eckart decomposition \cite{eckart}, and later make the due
corrections to take into account the one proposed by Landau and
Lifshitz \cite{Anderson, RelaBoltEqua}.

In the Eckart decomposition, the entropy 3-flow, defined by
\begin{equation}
  S^\alpha = -\int p^\alpha f \ln (f) \frac{d^2p}{p^0} \quad ,
\end{equation} 
can be written as follows:
\begin{equation}
  S^\alpha = n s U^\alpha + \varphi^\alpha \quad ,
\end{equation}
where $s = S^\alpha U_\alpha /n$ is the entropy per particle and
$\varphi^\alpha = \Delta_\beta^\alpha S^\beta$ the entropy flux.

In order to obtain the non-equilibrium distribution, we begin by
maximizing the entropy per particle, under the following constraints:
\begin{eqnarray}
  N^\alpha U_\alpha &= U_\alpha \int p^\alpha f \frac{d^2p}{p^0}
  \quad , \nonumber \\
  T^{\alpha \beta} U_\alpha &= U_\alpha \int p^\alpha p^\beta f \frac{d^2p}{p^0}
  \quad , \nonumber \\  
  T^{<\gamma \beta> \alpha} U_\alpha &= U_\alpha \int p^\alpha
  p^{<\gamma} p^{\beta >} f \frac{d^2p}{p^0}
  \quad ,
\end{eqnarray}
where,
\begin{equation}
  p^{<\alpha} p^{\beta >} = p^{(\alpha} p^{ \beta)} - \frac{1}{2}
  \Delta^{\alpha \beta} \Delta_{\gamma \delta} p^{(\gamma}p^{\delta)}
  \quad ,
\end{equation}
and 
\begin{equation}
  p^{(\alpha} p^{\beta )} = \frac{1}{2} ( \Delta^\alpha_\gamma
  \Delta^\beta_\delta + \Delta^\beta_\gamma \Delta^\alpha_\delta )
  p^{\gamma}p^{\delta} \quad .
\end{equation}

In principle, the moments $N^\alpha$ and $T^{\alpha \beta}$ would be a
more natural choice.  However, the resulting procedure to compute the
Lagrange multipliers via entropy maximization, while leading to equivalent 
results \cite{RelaBoltEqua}, proves significantly more complicated.
The problem of maximizing the entropy is equivalent to consider the
following functional,
\begin{equation}
  {\cal{F}} = s - \lambda N^\alpha U_\alpha - \lambda_\beta T^{\alpha
    \beta} U_\alpha - \lambda_{<\gamma \beta>} T^{<\gamma \beta>
    \alpha} U_\alpha \quad ,
\end{equation}
and applying the functional derivative $\delta {\cal{F}}/\delta f =
0$. Here, $\lambda$, $\lambda_\beta$, and $\lambda_{<\gamma \beta>}$
are Lagrange multiplier that we must determine. In two dimensions,
there are nine independent multipliers, since by definition
$\eta^{\gamma \beta} \lambda_{<\gamma \beta>} = 0$, which represents
an extra equation.

Following the procedure, as a result, one can approximate the
non-equilibrium distribution function by,
\begin{equation}\label{EqD}
  f \simeq f^{\rm eq} \left [ 1 - n \left ( \lambda + \lambda_\beta
      p^\beta + \lambda_{<\gamma \beta>} p^\gamma p^\beta \right ) \right ] \quad .
\end{equation}

By decomposing the Lagrange multipliers in the orthogonal basis in
space-time,
\begin{equation}
  \lambda_\beta   = \lambda' U_\beta + \lambda'_\gamma
  \Delta^\gamma_\beta \quad ,
\end{equation}
\begin{eqnarray}
  \lambda_{<\gamma \beta>}  &= \Lambda U_\beta U_\gamma + \frac{1}{2}
  \Lambda_\alpha (\Delta_\gamma^\alpha U_\beta + \Delta^\alpha_\beta
  U_\gamma ) \nonumber \\ &+ \Lambda_{\alpha \delta} \left (  \Delta_\gamma^\alpha
    \Delta_\beta^\delta - \frac{1}{2} \Delta^{\alpha \delta}
    \Delta_{\gamma \beta}  \right) \quad .
\end{eqnarray}
and inserting these variables into the equilibrium distribution,
Eq. (\ref{EqD}), we obtain
\begin{eqnarray}\label{EqD2}
  f &= f^{\rm eq} \bigg[ 1 - n \lambda - n \left( \lambda' U_\beta +
    \lambda'_\gamma \Delta^\gamma_\beta \right) p^\beta \nonumber \\ &- n
  \bigg( \Lambda U_\beta U_\gamma + \frac{1}{2} \Lambda_\alpha
  (\Delta_\gamma^\alpha U_\beta + \Delta^\alpha_\beta U_\gamma )
  \nonumber \\ &+ \Lambda_{\alpha \delta} \left ( \Delta_\gamma^\alpha
    \Delta_\beta^\delta - \frac{1}{2} \Delta^{\alpha \delta}
    \Delta_{\gamma \beta} \right) p^\gamma p^\beta \bigg ) \bigg
  ] \quad .
\end{eqnarray}

In the Grad method, we need to determine the value of the Lagrange
multipliers in terms of the macroscopic fields, $n$, $U^\alpha$, $T$,
$\omega$, $P^{<\alpha \beta>}$, and $q^\alpha$ (being the particle
density, macroscopic 3-velocity, temperature, dynamic pressure,
pressure deviator, and heat flux, respectively). The dynamic pressure
is defined by $\omega = - \mu \nabla_\alpha U^\alpha$, and the
pressure deviator by $P^{<\alpha \beta>} = 2 \eta \nabla^{<\alpha}
U^{\beta>}$, being $\mu$ and $\eta$ the bulk and shear viscosities
respectively. In this procedure, we also need the moments of the
equilibrium distribution function, which have been introduced in
Appendix \ref{moments5th}.

Let us impose that the actual distribution function carries the same
first moment as the equilibrium distribution, namely:
\begin{equation}
  N^\alpha = \int f p^\alpha \frac{d^2p}{p^0} = \int f^{\rm eq}
  p^\alpha \frac{d^2p}{p^0} \quad ,
\end{equation}
and apply the projectors, $U_\alpha$ and $\Delta_\alpha^\beta$, to
obtain respectively the first two equations for the Lagrange
multipliers,
\begin{eqnarray}\label{equationslagrange}
  -n^2 [\lambda + 2T (3 T \Lambda + \lambda')] &= 0 \quad , \nonumber \\
  n T \lambda'_{\gamma}\Delta^{\gamma \beta} + 3 n T^2
  \Lambda_\gamma \Delta^{\gamma \beta} &= 0 \quad .
\end{eqnarray}
In order to obtain the other equations, we calculate the
energy-momentum tensor,
\begin{equation}\label{energymomentum}
  T^{\alpha \beta} = \int f p^\alpha p^\beta\frac{d^2p}{p^0} \quad ,
\end{equation}
and introduce the projectors:
\begin{eqnarray}\label{projectors}
  \left ( \Delta_\alpha^\gamma \Delta^\delta_\beta - \frac{1}{2}
    \Delta^{\gamma \delta} \Delta_{\alpha \beta}  \right )  T^{\alpha \beta} &=
  P^{<\gamma \delta>} \quad , \nonumber \\
  \Delta_\alpha^\gamma U_\beta T^{\alpha \beta} &=
  q^{\gamma} \quad , \nonumber \\
  \Delta_{\alpha\beta} T^{\alpha \beta} &= -2(p + \omega) \quad ,
  \nonumber \\
  U_\alpha U_\beta T^{\alpha \beta} &= \epsilon \quad ,
\end{eqnarray}
where $\epsilon$ and $p$ are the energy density and hydrostatic
pressure, respectively. 

Thus, by inserting the distribution function in
Eq. (\ref{energymomentum}), and taking the projectors defined in
Eq. (\ref{projectors}), we obtain the following relations, 
\begin{eqnarray}\label{fourteenmoments}
  \omega &= -n^2 T ( \lambda  + 3T (4T \Lambda + \lambda') ) \quad ,
  \nonumber \\
  \epsilon &= 2nT -2 n^2 T ( \lambda  + 3T (4T \Lambda + \lambda') )
  \quad , \nonumber\\
  q^\gamma &= 3n^2T^2 \Delta^{\delta \gamma} \lambda'_\delta + 12
  n^2T^3 \Delta^{\delta \gamma} \Lambda_{\delta} \quad , \nonumber\\
  P^{<\gamma \delta>} &= - 6 n^2 T^3 \Lambda^{<\gamma \delta>} \quad .
\end{eqnarray}

Note that by imposing the state equation for the ultrarelativistic
system, $\epsilon = 2nT$, implies that that $\omega$ becomes zero.
This is equivalent to say that the bulk viscosity vanishes, like in
three dimensions.  In addition, this gives $\lambda = -T \lambda'$,
$\Lambda = -\lambda'/6T$, where $\lambda'$ can take any value. For
simplicity, we will set $\lambda'=0$. The arbitrariness on this
parameter is due to the fact that, in the ultrarelativistic regime
(virtually massless excitations), the number of particles density,
$n$, and the temperature of the system, are not independent, since $n
\sim T^2$ (e.g. gas of photons). In other words, the number of
particles density is fixed once the energy density $\epsilon$ of the
system is chosen, and therefore one Lagrange multiplier falls apart.

In order to obtain the other Lagrange multipliers, we solve the system
of algebraic equations, Eqs.~(\ref{equationslagrange}) and
(\ref{fourteenmoments}), to obtain:
\begin{eqnarray}
  \Delta^{\delta \gamma} \Lambda_\delta &= \frac{1}{3n^2T^3} q^\gamma
  \quad , \nonumber\\
  \Lambda^{<\gamma \delta>} &= -\frac{1}{6 n^2 T^3} p^{<\gamma \delta>}
  \quad , \nonumber\\
  \Delta^{\delta \gamma} \lambda'_\delta &= -\frac{1}{n^2T^2} q^\gamma
  \quad . 
\end{eqnarray}

Replacing these equations into the definition of the non-equilibrium
distribution, we obtain:
\begin{eqnarray}\label{EqD3}
    f = f^{\rm eq} \bigg[ 1 + \frac{q_\beta p^\beta}{nT^2} &-
    \frac{1}{3nT^3} q_\gamma U_\beta p^\gamma p^\beta +
    \frac{p_{<\gamma \beta>}}{6 n T^3} p^\gamma p^\beta \bigg ] .
\end{eqnarray}

This is the non-equilibrium distribution function for a
two-dimensional ultrarelativistic system, as expressed in terms of the
nine moments. The results described here can also be obtained by using
the so-called triangle scheme \cite{IlyaBook}, which is equivalent to
the Grad method. In order to calculate the explicit values of the heat
flux and the pressure deviator, we have to solve the Boltzmann
equation.  For the purpose of this study, we choose two approaches for
the collision operator, the first one proposed by Marle
\cite{MarleModel}, and the second one by Anderson and Witting
\cite{Anderson}.

\section{3th order moments of the Non-eq Distribution}

The third order moment of the distribution function can be calculated
as follows:
\begin{equation}
  T^{\alpha \beta \gamma} = \int f p^\alpha p^\beta p^\gamma
  \frac{d^2p}{p^0} \quad .
\end{equation}

By replacing Eq.~(\ref{EqD3}) into this equation and rising the
indexes for the nine fields, we obtain for the third order moment,
\begin{eqnarray}
  T^{\alpha \beta \gamma} &= T^{\alpha \beta \gamma}_{E} +
  \frac{q^\epsilon}{n T^2} \eta_{\beta \epsilon} T^{\alpha \beta
    \gamma \delta}_E \nonumber \\ &- \frac{q^\epsilon U^\lambda}{3 n T^3}
  \eta_{\delta \epsilon} \eta_{\sigma \lambda} T^{\alpha \beta
    \gamma \delta \sigma}_E+ \frac{P^{<\epsilon \lambda>}}{6 n T^3}
  \eta_{\delta \epsilon} \eta_{\sigma \lambda} T^{\alpha \beta
    \gamma \delta \sigma}_E \quad .
\end{eqnarray}

Note that for an accurate calculation of the third order moment of the
distribution function, knowledge up to the fifth order moment of the
equilibrium distribution (denoted by subindex $E$) is required.  The
moments of the equilibrium distribution, Eq.~(\ref{MJ:eq}), are
introduced in Appendix \ref{moments5th}. Thus, replacing the
respective moments of the equilibrium distribution, we obtain the
third order moment,
\begin{eqnarray}\label{3rdmoment}
  T^{\alpha \beta \gamma} &= 15 n T^2 U^\alpha U^\beta U^\gamma
  \nonumber \\
  &- 3 n T^2 ( \eta^{\alpha \beta} U^{\gamma} + \eta^{\alpha \gamma}
  U^{\beta} + \eta^{\beta \gamma} U^{\alpha} ) \nonumber \\ &- 2 T (
  \eta^{\alpha \beta} q^{\gamma} + \eta^{\alpha \gamma} q^{\beta} +
  \eta^{\beta \gamma} q^{\alpha} ) \nonumber \\ &+ 10 T ( U^{\alpha} U^{\beta}
  q^{\gamma} + U^{\alpha} U^{\gamma} q^{\beta} + U^{\beta}
  U^{\gamma} q^{\alpha} ) \nonumber \\ & + 5 T ( p^{<\alpha \beta>} U^{\gamma}
  + p^{<\alpha \gamma>} U^{\beta} + p^{<\beta \gamma>} U^{\alpha}
  )\quad .
\end{eqnarray}

With this expression at hand, we are ready to consider the Boltzmann
equation. For the case of the Marle model, we have all the needed
quantities in place. However, for the Anderson Witting, some
corrections are required, which we shall be introduced in
Sec.~\ref{AW3rd}.

\subsection{Marle Model}

For the case of the Boltzmann equation for the Marle model,
Eq.~(\ref{Boltzmann:eqM}), it is assumed that, 
\begin{eqnarray}\label{eckart:eq}
  \int f \frac{d^2p}{p^0} &= \int f^{\rm eq} \frac{d^2p}{p^0} = A =
  A_E \quad , \nonumber\\
  \int f p^\alpha \frac{d^2p}{p^0} &= \int f^{\rm eq} p^\alpha
  \frac{d^2p}{p^0} = N^\alpha = N_E^\alpha \quad . 
\end{eqnarray}

These conditions are satisfied in the Eckart decomposition
\cite{eckart}, as in the three dimensional case. The first order
moment of the distribution, $N^\alpha$, is defined by $n U^\alpha$,
the second moment is defined by
\begin{equation}
  T^{\alpha \beta} = P^{<\alpha \beta>} - p \Delta^{\alpha \beta} +
  (U^\alpha q^{\beta} + U^\beta q^{\alpha}) + \epsilon U^\alpha
  U^\beta \quad ,
\end{equation}
and the third order moment is calculated as described before, via
Eq.~(\ref{3rdmoment}). These are functions of $U^\alpha$, so that we
do not need any correction to the moment definitions.  By integrating
the Boltzmann equation, Eq.~(\ref{Boltzmann:eqM}), in the momentum
space, and taking into account the relations (\ref{eckart:eq}), we
obtain, $\partial_\alpha N^\alpha = 0$, and by multiplying the
equation by $p^\beta$ and repeating the same procedure, we further
obtain $\partial_\alpha T^{\alpha \beta} = 0$.  These are the
conservation equations for $N^\alpha$ and $T^{\alpha \beta}$.

However, by multiplying the equations by $p^\beta p^\gamma$, we obtain
a different equation, $\partial_\alpha T^{\alpha \beta \gamma} =
-(1/\tau_M) (T^{\beta \gamma} - T^{\beta \gamma}_E )$, which contains
the information about the transport coefficients.  By a standard
iteration procedure \cite{RelaBoltEqua}, we can convert this equation
into
\begin{equation}\label{marlediss}
  T^{\beta \gamma} - T^{\beta \gamma}_E = -\tau_M \partial_\alpha
  T^{\alpha \beta \gamma}_E \quad .
\end{equation}
This means that for the Marle model, we just need to know the third
order moment of the equilibrium distribution, and the second order
moment of both, the non-equilibrium and equilibrium distributions.
The corresponding transport coefficients will be calculated in
Sec.~\ref{tcoe}.

\subsection{Anderson-Witting Model}\label{AW3rd}

For the case of the Anderson-Witting model, we should use the
Landau-Lifshitz decomposition \cite{Anderson, RelaBoltEqua}.  Such
decomposition implies that $U^\alpha$ must be calculated, by solving
the eigenvalue problem, $T^{\alpha \beta}U_{L \alpha} = \epsilon U_{L
  \alpha}$ (subindex $L$ denotes Landau-Lifshitz). In general,
$U^\alpha$, calculated with the Eckart decomposition using $N^\alpha =
n U^\alpha$, will differ from the one calculated with the energy flux,
$U_{L}^\alpha$. As a consequence, we must find the relation between
both quantities and the correct expression for the third order kinetic
moment.

In the Landau-Lifshitz decomposition, we assume that $N^\alpha
U_\alpha = N^\alpha_E U_\alpha$, $U_\alpha T^{\alpha \beta} = U_\alpha
T^{\alpha \beta}_E$.  Moreover, the first and second order moments of
the distribution are defined by,
\begin{eqnarray}
  N^{\alpha} &= n U^\alpha_L + J^\alpha \quad , \nonumber \\
  T^{\alpha \beta} &= P_L^{<\alpha \beta>} - p_L \Delta_L^{\alpha \beta} + \epsilon_L U_L^\alpha
  U_L^\beta \quad . 
\end{eqnarray}

It can be easily shown that the correspondence between $U^\alpha$ and
$U^\alpha_L$ for the ultrarelativisitic case is $U^\alpha = U^\alpha_L
- q^\alpha/3 n$, with $J^\alpha = - q^\alpha/3 T$.  The conservation
equations $\partial_\alpha N^\alpha = 0$ and $\partial_\alpha
T^{\alpha \beta} = 0$ can be obtained by multiplying by $1$ and
$p^\beta$, and integrating the Boltzmann equation in the momentum
space, respectively.

By multiplying by $p^{\beta} p^{\gamma}$ and applying the Maxwell
iteration procedure as before, we obtain:
\begin{equation}\label{AWdiss}
  (T^{\alpha \beta \gamma} - T^{\alpha \beta \gamma}_E) U_{L\alpha} = -\tau \partial_\alpha
  T^{\alpha \beta \gamma}_E \quad .
\end{equation}
Note that in this case, at variance with the Marle case, the third
order moment of the non-equilibrium and equilibrium distribution
functions is needed.  To calculate the correct expression for the
third order moment, it is sufficient to replace $U^\alpha$ by
$U^\alpha_L - q^\alpha/3 n$ into Eq.~(\ref{3rdmoment}), retaining up
to linear terms in the nine fields.  This delivers:
\begin{eqnarray}\label{3rdmomentAW}
  T^{\alpha \beta \gamma} &= 15 n T^2 U_L^\alpha U_L^\beta U_L^\gamma \nonumber\\
  &- 3 n T^2 ( \eta^{\alpha \beta} U_L^{\gamma} + \eta^{\alpha \gamma}
  U_L^{\beta} + \eta^{\beta \gamma} U_L^{\alpha} ) \nonumber\\ &- T (
  \eta^{\alpha \beta} q^{\gamma} + \eta^{\alpha \gamma} q^{\beta} +
  \eta^{\beta \gamma} q^{\alpha} ) \nonumber\\ &+ 5 T ( U_L^{\alpha} U_L^{\beta}
  q^{\gamma} + U_L^{\alpha} U_L^{\gamma} q^{\beta} + U_L^{\beta}
  U_L^{\gamma} q^{\alpha} ) \nonumber \\ & + 5 T ( P^{<\alpha \beta>} U_L^{\gamma}
  + P^{<\alpha \gamma>} U_L^{\beta} + P^{<\beta \gamma>} U_L^{\alpha}
  )\quad .
\end{eqnarray}

Everything being in place, we next proceed to calculate the transport
coefficients for the two-dimensional ultrarelativistic system, using
both decompositions.

\section{Transport coefficients}\label{tcoe}

Since we found that the bulk viscosity vanishes, we focus on the shear
viscosity and the thermal conductivity.  First, let us consider the
Marle model and use the expressions (\ref{projectors}.  By applying
the projector $\Delta_\beta^\delta U_\gamma$ to Eq.~(\ref{marlediss}),
we obtain the heat flux,
\begin{equation}
  q^\delta = 3nT \tau_M \left ( \nabla^\delta T - \frac{1}{3n}
    \nabla^\delta p \right ) \quad ,
\end{equation}
and by applying the projector $\Delta_\beta^{(\epsilon} \Delta^{\delta
  )}_\gamma - \frac{1}{2} \Delta^{\epsilon \delta} \Delta_{\beta
  \gamma}$, we obtain the pressure deviator,
\begin{equation}
  P^{<\alpha \beta>} = 6 n T^2 \tau_M \nabla^{<\alpha} U^{\beta >} \quad .
\end{equation}
From this two relations we can conclude that the transport
coefficients in the model of Marle are given by,
\begin{equation}
  \kappa_M = \frac{ 3 c^2 k_B}{\xi} n \tau_M, \quad \eta_M = \frac{3}{\xi} n
  k_B T \tau_M, \quad \mu_M = 0
  \quad ,
\end{equation}
being the thermal conductivity, the shear viscosity, and the bulk
viscosity, respectively. Note that we have restablished the physical
units.

For the case of the Anderson-Witting model, we use Eq.~(\ref{AWdiss})
and apply the same projectors, finding
\begin{equation}
    q^\delta = \frac{3}{8} nT \tau \left ( \nabla^\delta T - \frac{1}{3n}
    \nabla^\delta p \right ) \quad ,
\end{equation}
for the heat flux and 
\begin{equation}
 P^{<\alpha \beta>} = \frac{6}{5} n T \tau \nabla^{<\alpha} U^{\beta >} \quad .
\end{equation}
for the pressure deviator. With these expressions, the transport
coefficients take the form:
\begin{equation}
  \kappa_{AW} = \frac{3 c^2 k_B}{8} n \tau, \quad \eta_{AW} =
  \frac{3}{5} n k_B T \tau, \quad \mu_{AW} = 0
  \quad .
\end{equation}

Here, as in the case of the Marle model, we have restored the physical
units. Note that the main difference between the transport
coefficients, apart from different numerical prefactors, is that the
ones calculated with the Anderson and Witting collision operator have
a different dependence on the temperature than the ones calculated
with the Marle model (since $\xi$ also depends on $T$).

\begin{figure}
\centering
\includegraphics[scale=0.37]{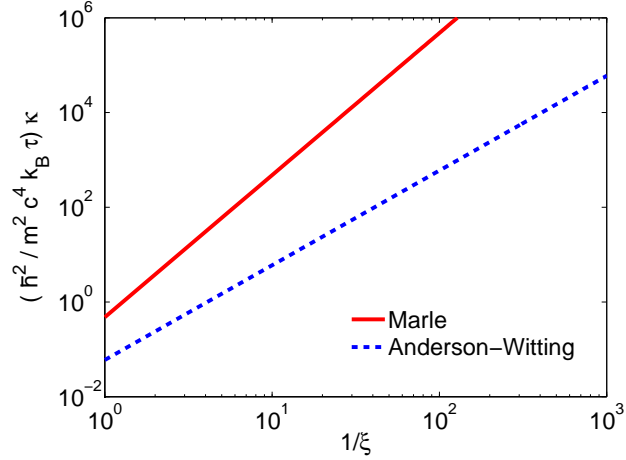}
\caption{Thermal conductivity $\kappa$ as a function of $\xi$. In this
  calculation we have set $\tau = \tau_M$.  The Marle coefficient is
  systematically higher than the Anderson-Witting one, and the ratio
  of the two grows at increasing $1/\xi$, i.e. at increasing
  temperature.}
\label{fig1}
\end{figure}

\begin{figure}
\centering
\includegraphics[scale=0.37]{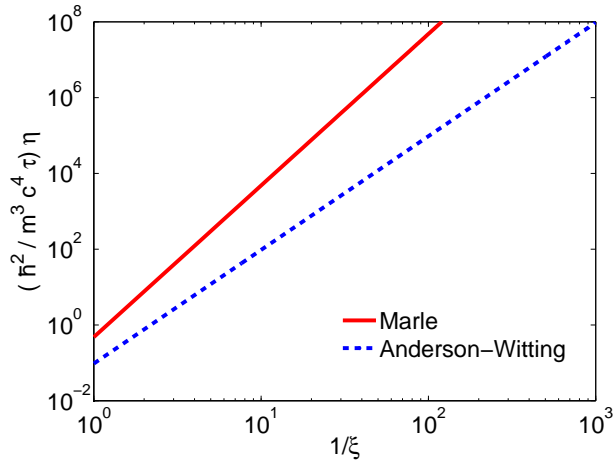}
\caption{Shear viscosity $\eta$ as a function of $\xi$. In this
  calculation we have set $\tau = \tau_M$.  Like for the case of
  Figure 1, the Marle coefficient is systematically higher than the
  Anderson-Witting one, and the ratio of the two grows at increasing
  $1/\xi$, i.e. at increasing temperature.  }
\label{fig2}
\end{figure}

Considering a non-degenerate gas of relativistic particles in the
ultrarelativistic regime, the number of particles density is given by
$n = k_B^2 T^2/ 2 \pi c^2 \hbar^2$. By taking this into account, we
see from Fig.~\ref{fig1} that the thermal conductivity $\kappa_M$
decreases with $\xi^3$, while $\kappa_{AW}$ with $\xi^2$. On the other
hand, in Fig.~\ref{fig2}, we can observe that the shear viscosity,
displays the same qualitative behavior, $\eta_M$ decreases with
$\xi^4$ while $\eta_{AW}$ with $\xi^3$. In general, given any
relativistic system, one can test which single relaxation time
approximation, Marle or Anderson-Witting, better reproduces its
behavior.  We have considered only values $1/\xi > 1$, since our
calculations are valid only in this regime.

An interesting calculation is to apply the corrections proposed by
Takamoto \cite{Takamoto} to the Marle model to account properly for
the ultrarelativistic regime, $\xi \rightarrow 0$, of a gas of
particles. Although this work was developed in $3+1$ dimensional
space-time, we will follow a similar procedure for the case of $2+1$
dimensions.  To this purpose, we replace the relaxation time $\tau_M$
by its average in momentum space, namely:
\begin{equation}
  \tau_M = \frac{mc}{n} \int \frac{d^2p}{p^0} f^{eq} \tau_{rel} =
  \xi \tau_{rel} \quad ,
\end{equation}
where $\tau_{rel}$ is now the effective relaxation time of the system
and $\tau_M$ a simple parameter in the relativistic Boltzmann
equation.  By replacing this relation in the equations for the
transport coefficients in the case of the Marle model, we obtain
\begin{equation}
  \kappa_{rel} = 3 c^2 k_B n \tau_{rel}, \quad \eta_{rel} = 3 n k_B T \tau_{rel}, \quad \mu_{rel} = 0
  \quad .
\end{equation}
Note that these transport coefficients carry the same dependence on
temperature as in the case of the Anderson-Witting model.  The
numerical coefficients, though, are not the same.


\section{Conclusions and Discussions}

In this work, we have calculated the transport coefficients, namely
the bulk and shear viscosities, and the thermal conductivity of a two
dimensional ultra-relativistic system, using two different forms of
the collision operator. The first one is based on the Marle model and
the second one on the Anderson Witting approach. Depending on the
approach, we have to satisfy the Eckart or the Landau-Lifshitz
decompositions, respectively.  This leads to different expressions for
the transport equations and third order moment of the distribution
function.

We have found that the bulk viscosity of the ultrarelativistic system
disappears as a consequence of the choice of the two-dimensional
ultra-relativistic equation of state, which imposes a constraint on
the trace of the momentum-energy tensor.  This is the same behavior
observed for the three dimensional case. By analyzing the transport
coefficients for the case of an ultrarelativistic gas of particles, we
have found that the thermal conductivity decreases with $\xi^3$ and
$\xi^2$, for the case of Marle and Anderson-Witting, respectively. The
shear viscosity presents the same qualitative behavior, decreasing
with $\xi^4$ and $\xi^3$ for both models, respectively. Therefore, the
Marle model transport coefficients always decrease faster than the
ones based on the Anderson-Witting model. In a more general
relativistic system, by knowing this difference, one could select
which one of the two is better suited to describe its dynamics
evolution. In addition, following the work by Takamoto
\cite{Takamoto}, we have modified the two-dimensional transport
coefficients for the case of the Marle model, in such a way as to make
it suitable for a gas of ultra-relativistic particles.  With such
modification, the functional dependence of the transport coefficients
on the temperature becomes the same as for the Anderson-Witting model,
although with different numerical coefficients.

It is known that transport coefficients in 2d are formally infrared
divergent, hence their size and gradient dependence must be taken with
some caution in practical applications
\cite{transanomalous,transanomalous2}. The investigation of these
issues in the relativistic framework makes an interesting object of
future research.

% {\color{red} Non-relativistic hydrodynamics in less than three
%   dimensions is affected by infinities, and cannot be linearized.
%   %However, the results in this paper are limited to the
%   %kinetic level, i.e. the Boltzmann equation, and therefore, the
%   %relativistic analogue of these divergences cannot be studied. 
% The divergence of the effective relaxation time in the Marle Model,
% $\tau = \tau_M E/mc^2$, is simply due to the fact that the factor
% $E/mc^2$ diverges in the limit $m \rightarrow 0$, for any finite
% value of the spatial momentum $|\vec{p}|$ (in the massless case, $E
% = c |\vec{p}|$).  This divergence is therefore completely unrelated
% to the divergence of transport coefficients in 2d, as due to
% non-Boltzmann effects (long-time tail of the velocity
% autocorrelation).  Such non-Boltzmann effects are ruled out by
% definition in our Boltzmann-based kinetic model.}
The results presented in this paper can be applied to a variety of
ultrarelativistic systems, e.g. graphene, plasma jets and others.  The
method is not limited to ultra-relativistic gases of particles, and it
extends to any statistical system obeying relativistic Boltzmann-like
equations.

\section*{Acknowledgments}
We acknowledge financial support from the European Research Council
(ERC) Advanced Grant 319968-FlowCCS. I.V.K. was supported by the ERC
Grant ELBM.

\appendix

\section{Moments of the Equilibrium Distribution}\label{moments5th}

The moments of the equilibrium distribution for a two-dimensional
ultrarelativistic system that satisfies the Maxwell J\"uttner
distribution are given by,
\begin{equation}
  N_E^\alpha = n U^\alpha \quad ,
\end{equation}
\begin{equation}
  T_E^{\alpha \beta} = - n T \eta^{\alpha \beta} + 3 n T U^\alpha U^\beta \quad ,
\end{equation}
\begin{eqnarray}
  T_E^{\alpha \beta \gamma} = - 3 n T^2 (\eta^{\alpha \beta}
  U^\gamma &+ \eta^{\alpha \gamma} U^\beta + \eta^{\beta \gamma}
  U^\alpha) \nonumber\\ &+ 15 n T^2 U^\alpha U^\beta U^\gamma \quad ,
\end{eqnarray}
\begin{eqnarray}
  T_E^{\alpha \beta \gamma \delta} &= 3 n T^3 (\eta^{\alpha \beta}
  \eta^{\gamma \delta} + \eta^{\alpha \gamma} \eta^{\beta \delta}
  + \eta^{\gamma \beta} \eta^{\alpha \delta} ) \nonumber\\ &- 15 n T^3 (
  \eta^{\alpha \beta} U^\gamma U^\delta + \eta^{\alpha \gamma}
  U^\beta U^\delta + \eta^{\gamma \beta} U^\alpha U^\delta \nonumber\\ &+
  \eta^{\alpha \delta} U^\gamma U^\beta + \eta^{\delta \gamma}
  U^\beta U^\alpha + \eta^{\delta \beta} U^\alpha U^\gamma ) \nonumber\\ &+ 105
  n T^3 U^\alpha U^\beta U^\gamma U^\delta \quad ,
\end{eqnarray}
\begin{eqnarray}
  T_E^{\alpha \beta \gamma \delta \epsilon} &= 15 n T^4 [
  U^\epsilon (\eta^{\alpha \beta} \eta^{\gamma \delta} +
  \eta^{\alpha \gamma} \eta^{\beta \delta} + \eta^{\gamma \beta}
  \eta^{\alpha \delta} ) \nonumber\\ &+ U^\alpha (\eta^{\epsilon \beta}
  \eta^{\gamma \delta} + \eta^{\epsilon \gamma} \eta^{\beta
    \delta} + \eta^{\gamma \beta} \eta^{\epsilon \delta} ) \nonumber\\ &+
  U^\beta (\eta^{\alpha \epsilon} \eta^{\gamma \delta} +
  \eta^{\alpha \gamma} \eta^{\epsilon \delta} + \eta^{\gamma
    \epsilon} \eta^{\alpha \delta} ) \nonumber\\ &+ U^\beta (\eta^{\alpha
    \epsilon} \eta^{\gamma \delta} + \eta^{\alpha \gamma}
  \eta^{\epsilon \delta} + \eta^{\gamma \epsilon} \eta^{\alpha
    \delta} ) \nonumber\\ &+ U^\gamma (\eta^{\alpha \beta} \eta^{\epsilon
    \delta} + \eta^{\alpha \epsilon} \eta^{\beta \delta} +
  \eta^{\epsilon \beta} \eta^{\alpha \delta} ) \nonumber\\ &+ U^\delta
  (\eta^{\alpha \beta} \eta^{\gamma \epsilon} + \eta^{\alpha
    \gamma} \eta^{\beta \epsilon} + \eta^{\gamma \beta}
  \eta^{\alpha \epsilon} ) ] \nonumber\\ & - 105 n T^4 ( \eta^{\alpha
    \beta} U^\gamma U^\delta U^\epsilon + \eta^{\alpha \gamma}
  U^\beta U^\delta U^\epsilon \nonumber\\ &+ \eta^{\alpha \delta} U^\beta
  U^\gamma U^\epsilon + \eta^{\delta \gamma} U^\alpha U^\beta
  U^\epsilon \nonumber\\ &+ \eta^{\delta \beta} U^\alpha U^\gamma
  U^\epsilon + \eta^{\beta \gamma} U^\alpha U^\delta U^\epsilon \nonumber\\
  &+ \eta^{\alpha \epsilon} U^\beta U^\gamma U^\delta +
  \eta^{\beta \epsilon} U^\alpha U^\gamma U^\delta + \eta^{\gamma
    \epsilon} U^\alpha U^\beta U^\delta ) \nonumber\\ &+ 945 n T^4 U^\alpha
  U^\beta U^\gamma U^\delta U^\epsilon \quad .
\end{eqnarray}

To obtain the moments, we have considered the ultrarelativistic regime, $\xi \ll 1$.

\section*{References}
\bibliographystyle{unsrt}
\bibliography{report}

\end{document}